\documentclass[conference,a4paper]{IEEEtran}
\ifCLASSINFOpdf
\else
\fi
\usepackage{amsmath}
\usepackage{graphicx}
\usepackage{latexsym,amssymb,epsf}

\begin{document}
\title{Partially Coherent Optical Modelling of the Ultra-Low-Noise Far-Infrared
Imaging Arrays on the SPICA Mission}

\author{\IEEEauthorblockN{Stafford Withington}
\IEEEauthorblockA{Cavendish Laboratory\\
J.J. Thomson Avenue\\
Cambridge CB3 OHE\\
Email: stafford@mrao.cam.ac.uk}
\and
\IEEEauthorblockN{Christopher N. Thomas}
\IEEEauthorblockA{Cavendish Laboratory\\
J.J. Thomson Avenue\\
Cambridge CB3 OHE\\
Email: cnt22@cam.ac.uk}
\and
\IEEEauthorblockN{David J. Goldie}
\IEEEauthorblockA{Cavendish Laboratory\\
J.J. Thomson Avenue\\
Cambridge CB3 OHE\\
Email: goldie@mrao.cam.ac.uk}}


\maketitle

\begin{abstract}
%\boldmath
We have developed a range of theoretical and numerical techniques for modeling the multi-mode, 210-34~$\mu$m, ultra-low-noise Transition Edge Sensors that will be used on the SAFARI instrument on the ESA/JAXA cooled-aperture FIR space telescope SPICA. The models include a detailed analysis of the resistive and reactive properties of thin superconducting absorbing films, and a partially coherent mode-matching analysis of patterned films in multi-mode waveguide. The technique allows the natural optical modes, modal responsivities, and Stokes maps of complicated structures comprising patterned films in profiled waveguides and cavities to be determined.
\end{abstract}

{\smallskip \keywords FIR detectors, Transition Edge Sensors, multi-mode bolometers, mode matching,
superconducting absorbers}

% \fi

\IEEEpeerreviewmaketitle

\section{Introduction}

SPICA is a European Space Agency (ESA), Japanese Space Agency (JAXA) mission to place a cooled-aperture far-infrared space telescope at Lagrange point L2. The 3.2~m primary mirror will be cooled to 5~K, giving extraordinary levels of performance. We are working with the Space Research Organisation of the Netherlands (SRON), the University of Cardiff UK, and the National University of Ireland Maynooth, to develop the ultra-low-noise, 210-110~$\mu$m (324 pixel), 110-60~$\mu$m (1156 pixel), and 60-34~$\mu$m (1849 pixel), imaging arrays needed for this mission. The arrays require pioneering levels of performance, \cite{Jackson_1} and in this context we have demonstrated Transition Edge Sensors (TES), operating at 70mK, achieving NEP's of 4$\times$10$^{-19}$~WHz$^{-1/2}$.

Each pixel comprises a tiny far-infrared (FIR) lightpipe that is coupled to a suspended superconducting absorber, which is thermally linked to a MoAu or TiAu Transition Edge Sensor (TES). Many similar pixels are packed together to form large-format focal plane imaging arrays. The optical design is unconventional because each detector is sensitive to a small number, typically 20, electromagnetic modes, which contrasts with microwave horns, which are sensitive to a single mode, and large planar optical pixels, which are sensitive to thousands of modes.

As part of our development programme we have devised a variety of partially coherent electromagnetic modeling techniques.  A crucial point is that each detector is sensitive to a number of optical modes, and we wish to determine the amplitude, phase, and polarisation patterns of each of these modes, and their individual responsivities. These modes must be propagated through the full optical system of the telescope to determine overall performance. The SAFARI detectors saturate with very low levels of optical power, 20~fW, and so it is essential to understand their sensitivity to stray light.

\section{Superconducting absorbers}

In the SAFARI detectors, the FIR absorber comprises a thin ($<$10~nm) $\beta$-phase superconducting tantalum (Ta) absorber on a 200~nm thick SiN membrane. The Ta is superconducting ($T_{c} =$ 860~mK) at the operating temperature ($<$100~mK), and so its heat capacity is small, allowing fast response times (10~ms) to be achieved. The FIR photons, however, have enough energy to break Cooper pairs, and so the Ta film is electromagnetically absorptive. It seems reasonable that a thin superconducting film can be modeled as a parallel sheet resistance, but it is important that the sheet-impedance model works both with regards to absorption and scattering, as a long electromagnetic coherence length would reduce the modal throughput of the absorber, and the complete instrument.
\begin{figure}
\centering
\includegraphics[width=2.4in]{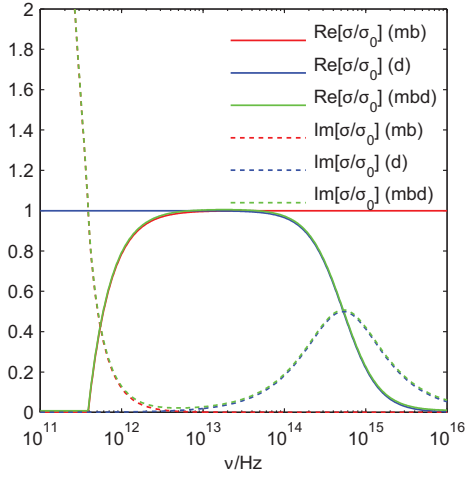}
\caption{Normalized conductivity of Ta at 100~mK based on Mattis Bardeen (red), AC Drude (blue), and a combined Mattis-Bardeen-Drude model (green). The solid lines show the real parts and the dashed lines the imaginary parts.}
\label{fig_1}
\end{figure}

To construct a model of a few-mode TES detector, it is first necessary to understand the intrinsic spectral form of the complex conductivity of the superconducting absorber. We used a composite model of bulk conductivity $\sigma(\omega)$ based on the local Mattis-Bardeen \cite{Mattis_1} and Drude \cite{Drude_1} approaches. The Mattis-Bardeen model includes both thermally and optically excited quasiparticles, as well as the kinetic inductance effect of Cooper pairs, whereas the Drude model characterizes the onset of UV transparency.

Fig.~\ref{fig_1} shows the complex conductivity of ordinary Ta normalized to the measured, normal-state, DC conductivity of $\beta$-phase Ta, $\sigma_{0} = 5 \times 10^{5}$~Sm$^{-1}$, to give the correct sheet impedances. The superconducting properties of ordinary Ta are different to those of $\beta$-phase Ta, but the analysis is sufficient for our purposes. The conductivity is largely frequency independent and resistive over the SAFARI operating range, 1.4 to 8.8~THz. The increasingly reactive behaviour below 1~THz is caused by the blocking of final states by condensed quasiparticles as the gap frequency (380~GHz for Ta, 62~GHz for $\beta$-Ta) is approached. At the highest frequency 8.8~THz, the skin depth is of order 250~nm, the film thickness is less than 10~nm, and the wavelength in the film is 1.5~$\mu$m, indicating that the film is much thinner than the electromagnetic coherence length, and that a fully coherent electromagnetic representation is suitable for describing the penetration of the field into the absorber, validating a sheet-impedance model.

\begin{figure}
\centering
\includegraphics[width=2.4in]{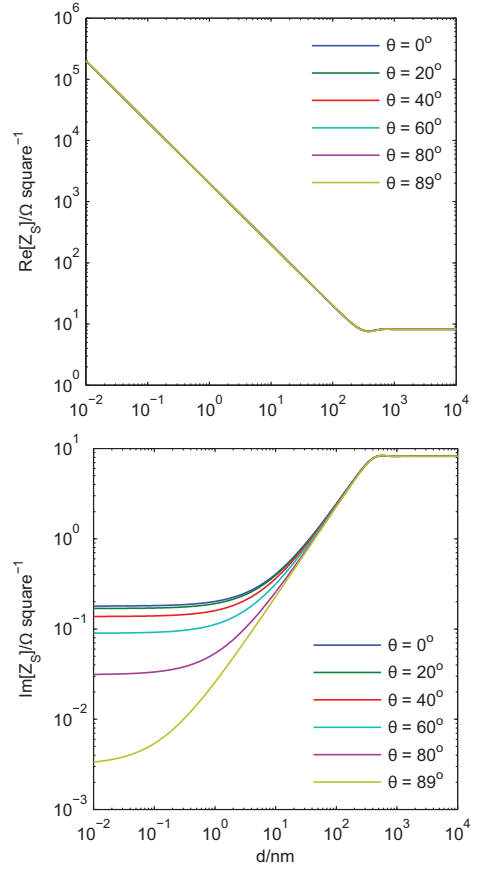}
\caption{Real (top) and imaginary (bottom) parts of the sheet impedance of a superconducting Ta film at a wavelength of 35~$\mu$m for s-polarized plane-wave excitation at various angles of incidence $\theta$.}
\label{fig_2}
\end{figure}

\begin{figure}
\centering
\includegraphics[width=2.4in]{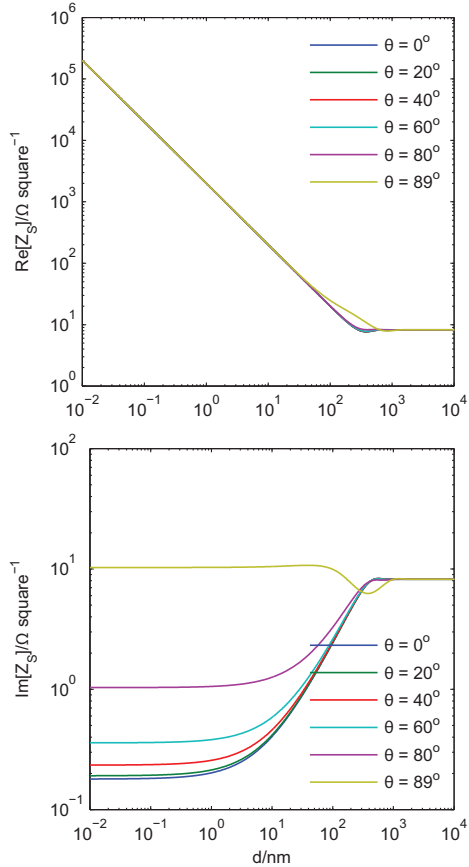}
\caption{Real (top) and imaginary (bottom) parts of the sheet impedance of a superconducting Ta film at a wavelength of 35~$\mu$m for p-polarized plane-wave excitation at various angles of incidence $\theta$.}
\label{fig_3}
\end{figure}

Using the Mattis-Bardeen-Drude conductivity, we have calculated the sheet impedance in two different ways: (i) a plane-wave method, and (ii) a Green's dyadic method; both gave identical results. In (i) the incident and transmitted fields take the form of plane waves, and the field within the conductor is also described in terms of plane waves. The fields on either side of the front and back surfaces are matched in the usual way. By calculating the current distribution in the film, the sheet impedance can be determined \cite{Thomas_1}.

Fig.~\ref{fig_2} shows the real (top) and imaginary (bottom) parts of the sheet impedance, at a wavelength of 35~$\mu$m, as a function of film thickness for s-polarized plane waves (normal to the plane of incidence). Fig.~\ref{fig_3} shows the results for p-polarized plane waves (parallel to the plane of incidence). The real parts of the sheet impedances are largely independent of angle of incidence and polarization, and scale with film thickness $d$ for thicknesses below the penetration depth $\delta$, (250~nm): ${\rm Re} [Z_{s}] \approx ( \sigma d)^{-1}$. For thicknesses above the penetration depth, the calculated impedance becomes that of the surface impedance of an infinite half space: ${\rm Re} [Z_{s}] \approx ( \sigma \delta)^{-1}$. In order to achieve a sheet impedance of 377~$\Omega$~square$^{-1}$, $d\approx$~4~nm is needed. Sub-wavelength patterning can be used to increase the sheet impedance without reducing film thickness.
\begin{figure}
\centering
\includegraphics[width=2.4in]{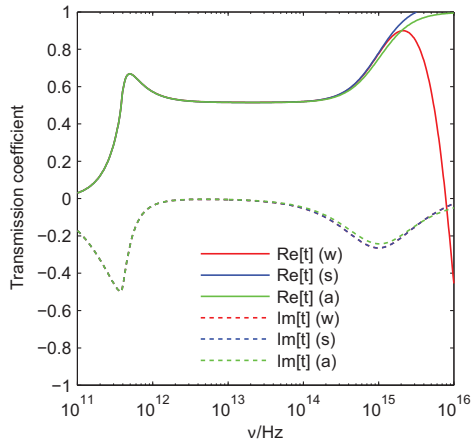}
\caption{Real and imaginary parts of the electric-field transmission coefficient of a 10~nm film of $\beta$-phase Ta at 100~mK and normal incidence. The different lines correspond to the different models described in the text.}
\label{fig_4}
\end{figure}
\begin{figure}
\centering
\includegraphics[width=2.4in]{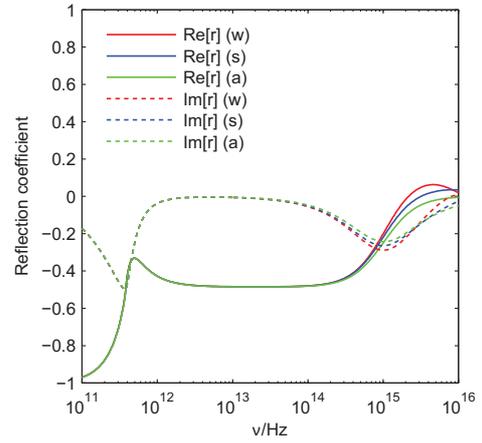}
\caption{Real and imaginary parts of the electric-field reflection coefficient of a 10~nm film of $\beta$-phase Ta at 100~mK and normal incidence. The different lines correspond to the different models described in the text.}
\label{fig_5}
\end{figure}

The sheet reactances have very different behaviour in the s- and p-polarized cases, with the highest reactances appearing for p-polarized waves. For moderate angles of incidence, and film thicknesses of $<$10~nm, the reactive part is almost three orders of magnitude smaller than the resistive part, and frequency-independent absorption is possible.
This situation occurs because although the evanescent waves in the film have real and imaginary parts, the real parts of the forward and backward waves add, whereas the imaginary parts cancel. This explains why the sheet impedance of a thin film is real, whereas the surface impedance of a bulk conductor is complex.

Figures \ref{fig_4} and \ref{fig_5} show the electric-field transmission and reflection coefficients of a 10~nm sheet (chosen for impedance match when the film is free-standing) of Ta at 100~mK as a function of frequency for various models: (w) full plane-wave decomposition; (s) plane-wave model assuming that the film is thinner than a skin depth, but taking into account reactive behaviour; (a) a calculation based on $Z_{s} \approx ( \sigma d)^{-1}$. Over the SAFARI frequency range, 50~\% of the incident power is absorbed with 25~\% being reflected and 25~\% transmitted, which indicates near ideal absorptive behaviour despite the film being superconducting. The models only deviate at 10$^{15}$~Hz, where the skin depth starts to become comparable with film thickness.

\section{Mode matching}

A SAFARI detector comprises a sheet of superconducting absorber in a cavity at the back of a few-mode waveguide. Many questions arise: What shape of cavity should be chosen? What are the natural optical modes of the assembly and what are their individual responsivities? We have developed a rigorous mode-matching technique for analyzing the behavior of multiple layers of patterned films in profiled few-mode waveguides. The basic approach is to calculate the multimode scattering parameters of waveguide-mounted films, which can have any spatially varying complex conductivity, and therefore may include various materials: for example, taking into account the parasitic optical effects of the TES and wiring. By patterning the absorber it should be possible to minimize heat capacity, and even tailor the frequency response.

There are two approaches to calculating the scattering parameters of patterned films in multimode waveguide: (i) a Green's Dyadic Method (GDM), \cite{Withington_1} which uses a MoM-like approach in a waveguide environment; (ii) Boundary Matching Method (BMM) \cite{Murphy_1} where the jump conditions associated with the sheet impedance are used to match the waveguide mode coefficients on either side of a film. It can be proven analytically that these two methods are identical \cite{Thomas_1}.

Once the scattering parameters of the transverse sheet are known they can be cascaded with the scattering parameters of step changes in waveguide size to build up components having complex profiles. In addition, they can be cascaded with other thin-film structures to realize integrated components, such as metal-mesh filters in the front of the absorber. A feature of our approach is that we have derived expressions for the scattering parameters of an element where a patterned film is precisely at the same position as a step change in waveguide size. In this way, a single library element can be used to build up multilayer components having almost any waveguide profile.

Once the scattering matrices associated with films and steps have been calculated they can be cascaded into a single scattering matrix that fully characterizes the partially coherent optical behavior of the complete system. In the case of a detector, the scattering matrix ${\bf S}$ is simply the multimode reflection matrix looking into the aperture of the input waveguide. The question then arises as to how the overall efficiency and the natural modes responsible for power absorption can be found.

We do not formulate the problem in terms of the usual normalized mode coefficients, where the modulus-squared gives the power flow, but in terms of ordinary mode coefficients. As a consequence, our scattering parameters are not the usual microwave scattering parameters. The reason is that we wish to take into account evanescent modes, for which the real part of the mode impedance is zero and for which the normalized mode coefficients are undefined. It can be shown that the power $P$ absorbed by a waveguide detector is then given by
\begin{equation}
\label{1}
P = {\bf a}^{\dagger} \cdot {\bf D}^{\dagger} \cdot {\bf a}
\mbox{,}
\end{equation}
where ${\bf a}$ is the vector of mode coefficients at the input reference plane. If ${\bf Z}$ is the diagonal, non-Hermitian matrix of waveguide mode impedances, then the Hermitian response matrix
\begin{eqnarray}
\label{2}
{\bf D} & = & \frac{1}{2} \left( {\rm Re} \left[ {\bf Z}^{-1} \right] - {\bf S}^{\dagger} \cdot {\rm Re} \left[ {\bf Z}^{-1} \right] \cdot {\bf S} \right) \\ \nonumber
 & + & \frac{i}{2} \left( {\bf S}^{\dagger} \cdot {\rm Im} \left[ {\bf Z}^{-1} \right] - {\rm Im} \left[ {\bf Z}^{-1} \right] \cdot {\bf S} \right)
\end{eqnarray}
describes the ability of the system to absorb power. (\ref{2}) has two terms. The first describes power absorption through propagating modes. The second, which is conventionally not included, describes power absorption through evanescent modes. For example, certain modes will be close to cut off, and have long decay lengths inside the waveguide. A conducting film, close to the input aperture, may interact with these evanescent modes and absorb power. Although this effect is unlikely to be significant in practice, its inclusion ensures rigor.

Take the ensemble average of (\ref{1}) and express it in the form
\begin{equation}
\label{3}
\langle P \rangle = \mbox{Tr} \left[ \langle {\bf a} {\bf a}^{\dagger} \rangle \cdot {\bf D}^{\dagger} \right] =
\mbox{Tr} \left[ \bf{A} \cdot {\bf D}^{\dagger} \right]
\mbox{,}
\end{equation}
where Tr denotes the trace, and ${\bf A}$ is the coherence matrix of the incident modes. ${\bf A}$ describes the state of coherence of the incoming field. The Trace of the product of two Hermitian matrices is an inner product in the abstract space of matrices. (\ref{3}) describes the projection of the state of coherence of the incoming field onto the state of coherence to which the detector is sensitive. It is possible to diagonalize ${\bf D}$: the eigenvectors give the natural modes to which the partially coherent detector is sensitive, and the eigenvalues the responsivities in each of the natural modes. The Stokes fields can also be calculated.

\begin{figure}
\centering
\includegraphics[width=3.3in]{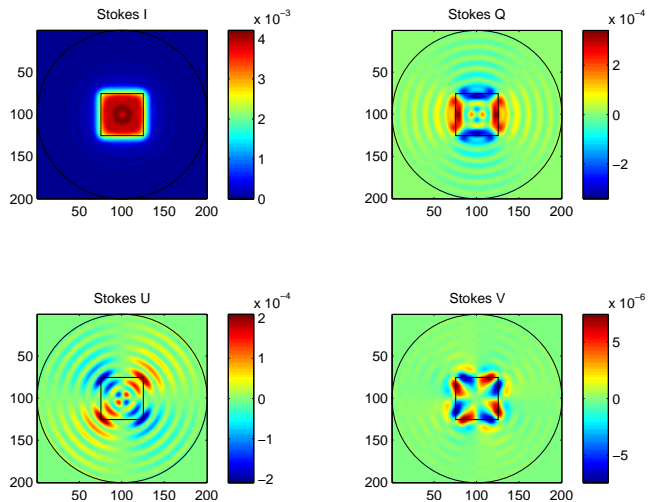}
\caption{Stokes maps of the partially coherent reception field of a square absorber in circular waveguide. The waveguide had a radius of 4~$\lambda$, the film had a side length of 2~$\lambda$, and the sheet impedance was 188.5~$\Omega$~square$^{-1}$. The black circle and black square indicate the physical sizes of the waveguide and absorber respectively.}
\label{fig_6}
\end{figure}

Fig.~\ref{fig_6} shows the Stokes reception maps (I,Q,U,V) looking into a multi-mode circular waveguide supporting a square absorber. The waveguide had a radius of 4~$\lambda$, the film had a side length of 2~$\lambda$, and the sheet impedance was 188.5~$\Omega$~square$^{-1}$. The output waveguide behind the absorber was assumed perfectly matched, but in the case of a real detector it would be shorted to form a perfectly reflecting boundary. The intensity $I$ plot shows that the edges and corners of the absorber are not well defined, and the $Q,U,V$ plots show that cross-polar response occurs as a result of currents flowing around the edges of the film. The forms of the natural modes are determined both by the finite size of the square absorber and the walls of the circular waveguide.

Although not shown here, in the extreme case when the absorber is much smaller than the waveguide and wavelength, the modal throughput is determined solely by the absorber. The modal response then comprises the two electric dipole modes and the one magnetic dipole mode of a sub-wavelength free-space pixel. \cite{Thomas_2} As the pixel is made larger, but still smaller than the waveguide, the modal forms and throughput are determined by the size and shape of both the absorber and waveguide. As the pixel is made larger still, until it completely fills the waveguide, the throughput is determined solely by the propagating waveguide modes. The natural modes become those of the propagating waveguide modes, with the individual efficiencies depending on the relative values of the modal waveguide impedances and the film sheet impedance. When a backshort is present, not all of the modes can be matched to the film because of the differing guide wavelengths, particularly near cut off. In the case of a film that spans the waveguide, a simple model based on an incoherent superposition of imperfectly matched waveguide modes is accurate.

\section{SAFARI detectors}

Originally, our SAFARI designs comprised a film in a hemispherical integrating cavity at the back of a multi-mode waveguide. Now, after many simulations, we have largely eliminated the role of the integrating cavity. The absorber is placed very close to the flat face of the cavity, such that the field diffracting away from the incoming waveguide interacts almost immediately with the absorber. A backshort, comprising a Au film on a micromachined Si pillar, is then placed within a fraction of a wavelength of the back of the film. The new arrangement has many advantages: (i) The dependence on cavity modes is eliminated, leading to wideband $>$50\% optical performance with little in-band structure. (ii) There are no currents in the walls of the rough mechanically machined cavity, which result in parasitic ohmic loss. The only reflective currents flow in a sputtered Au film on a polished Si surface. (iii) The absorber is kept as small as possible, which has important consequences for detector sensitivity and speed. The inter-pixel cross-talk is kept small by not cutting currents that would otherwise flow in the walls of the cavity. (iv) In the case of a multimode detector, the Stokes fields on the sky can be highly dependent on the precise shape of the cavity. Therefore eliminating the cavity will help minimize pixel-to-pixel variations in the optical response patterns.
\begin{figure}
\centering
\includegraphics[width=2.5in]{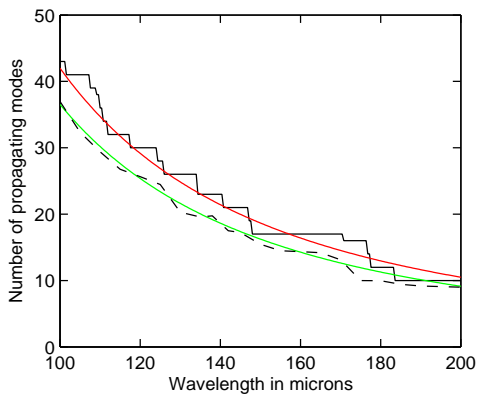}
\caption{Modal throughput of a film having a sheet impedance of 377~$\Omega$~square$^{-1}$ in a circular waveguide having a radius of 160~$\mu$m. A flat backshort was placed at 56~$\mu$m. The various models are described in the text.}
\label{fig_7}
\end{figure}

Various interesting effects become apparent: (i) The optimum distance needed to maximize overall efficiency is 0.375~$\lambda$, and not the 0.25~$\lambda$ that might be expected. The reason is that for any multi-mode system, there are always some modes close to cut off. As TE modes approach cut off their impedances become very high, whereas as TM modes become low. Due to their low impedances, TM modes close to cut off cause a particularly significant degradation in efficiency. By placing the backshort at 0.375~$\lambda$ one degrades the coupling efficiencies of modes well away from cut off, but increases the efficiencies of modes close to cut off, leading to the best overall behaviour. (ii) A highly optimized design never achieves greater than 87\% efficiency. The issue seems to relate to the fact that the currents that dissipate energy are two-dimensional. A volumetric absorber is capable of absorbing the power in all of the waveguide modes perfectly, but the currents in a planar absorber have fewer degrees of freedom.

Fig.~\ref{fig_7} shows the throughput, number of modes, of a long-wavelength 210-110~$\mu$m detector as a function of wavelength. Circular waveguide was used having a radius of 160~$\mu$m; the absorber had a nominal sheet impedance of 377~$\Omega$~square$^{-1}$ and filled the waveguide; a backshort was placed at 56~$\mu$m. We have plotted throughput rather than efficiency, because in an astronomical application, it is the throughput that is important. The models relate to (i) throughput based on a strict geometric notion of how many modes propagate at a given wavelength, (ii) a $1/ \lambda^{2}$ fit, (iii) the  $1/ \lambda^{2}$ fit scaled by 87~\%, and (iv) a full electromagnetic mode-matching simulation. The throughput of the full calculation changes relatively smoothly compared with the strict geometrical calculation, because waveguide modes do not cut on and off abruptly. The hazard of using efficiency relative to the number of propagating modes is evident, as the efficiency can seem to change abruptly depending on how close the wavelength is to a step change in the number of modes calculated geometrically. At certain wavelengths, 145~$\mu$m, the efficiency can seem exceedingly high, whereas at a neighboring wavelength, 140~$\mu$m, it is low. An average `efficiency' of 87\% is the maximum that can be achieved. A superconducting film with a flat backshort at 0.375~$\lambda$ can achieve near-ideal behaviour across a bandwidth of $>$60~\%.

\section{Conclusions}

We are using mode-matching software to design the ultra-low-noise multi-mode superconducting detectors for SAFARI. In particular, we recommend designs based on micromachined flat backshorts. Many enhancements are possible: (i) The absorbing film and backshort may be patterned to allow lower impedance $\beta$-phase Ta films to be used, which would lessen the demands and hazards of using films that are only a few nanometres thick. (ii) The absorbing film may be patterned to incorporate some level of filtering, if only to minimize the optical stray-light cross section. (iii) It may be possible to use a superconductng film on the backshort to help provide low-frequency magnetic shielding of the TES whilst still retaining a highly reflective optical coating. (iv) We would like to include a Au thermalizing ring around the outside of the absorber to help thermalization and minimize thermal fluctuation noise. This addition seems reasonable because the absorber is very close to the flat front wall of the hemispherical cavity and absorptive optical currents cannot flow in the part of the absorber that extends beyond the waveguide aperture.

\section*{Acknowledgements}

We would like to thank ESA for its support of the European ultra-low-noise FIR TES development programme.

\end{document}